\documentclass[aps,prb,reprint,superscriptaddress,floatfix]{revtex4-1}

\usepackage{graphicx}
\usepackage{color}

\begin{document}


\title{Quantum-Confined Stark Effect in polar and nonpolar  Wurtzite InN/GaN Heterostructures: Influence on Electronic Structure and Compensation by Coulomb Attraction}


\author{Stefan Barthel}
\email{sbarthel@itp.uni-bremen.de}
\author{Kolja Schuh}
\affiliation{Institute for Theoretical Physics, University of Bremen, Germany}

\author{Oliver Marquardt}
\affiliation{now at: Paul-Drude-Institut f\"ur Festk\"orperelektronik, Berlin, Germany}
\affiliation{Max-Planck Institut f\"ur Eisenforschung, D\"usseldorf, Germany}

\author{Tilmann Hickel}
\author{J\"org Neugebauer}
\affiliation{Max-Planck Institut f\"ur Eisenforschung, D\"usseldorf, Germany}

\author{Frank Jahnke}
\author{Gerd Czycholl}
\affiliation{Institute for Theoretical Physics, University of Bremen, Germany}


\date{\today}

\begin{abstract}
In this paper we systematically analyze the electronic structures of polar and nonpolar wurtzite-InN/GaN quantum dots and their modification due to the quantum-confined Stark effect caused by intrinsic fields. This is achieved by combining continuum elasticity theory with an empirical tight binding model to describe the elastic and single-particle electronic properties in these nitride systems. Based on these results, a many-body treatment is used to determine optical absorption spectra. The efficiency of optical transitions depends on the interplay between the Coulomb interaction and the quantum-confined Stark effect. We introduce an effective confinement potential which represents the electronic structure under the influence of the intrinsic polarization fields and calculate the needed strength of Coulomb interaction to diminish the separation of electrons and holes. 
\end{abstract}


\maketitle

\section{Introduction}
The quantum-confined Stark effect (QCSE) is the main reason for a reduced optical recombination efficiency in wurtzite nitride heterostructures grown along a polar axis. For instance, quantum dots will exhibit a charge carrier separation, see e.g. Ref.(\onlinecite{marquardt:083707}), since strong intrinsic electrostatic fields caused by piezo- and pyroelectric contributions drastically modify the confinement potential for electrons and holes. Motivated by the fact that the alloyed nitride systems can principally cover the whole visible range of emission wavelengths, one possible idea to overcome the drawback of the charge separation is to perform growth along a nonpolar facet, Refs.(\onlinecite{yang:061914,Wetzel20083987,founta:171901}). As a result, the spontaneous polarization part of the intrinsic fields can be reduced, which should lead to weaker spatial separation of electron and hole wave functions and thus to more efficient light emission. Driven by the large potential as an application in optoelectronic devices, see Ref.(\onlinecite{Nakamura14081998}), we theoretically analyze the electronic structure of polar (c-plane) and nonpolar (m-plane) InN/GaN quantum dots (QDs) under the influence of the persistent intrinsic electrostatic fields. To establish a more general understanding, we not only discuss ground state properties which have been studied together with the intrinsic fields in Refs.(\onlinecite{PhysRevB.79.081401,marquardt:083707,PSSC:PSSC200982609}), but also excited states of the system. These excited states play a major role if one includes Coulomb interaction and goes from a single-particle description to a many-body theory, which enables to study the competition between intrinsic fields and electron-hole attraction. The authors have already shown in Ref.(\onlinecite{schuh:092103}), that the Coulomb interaction does enhance the optical dipole strength in pure lenshaped nonpolar InN/GaN QDs. Recently, Schulz \textit{et al.} have studied nonpolar GaN/AlN QDs by taking into account the precise QD geometry and using a self-consistent Hartree approximation to account for Coulomb effects in Ref. (\onlinecite{schulz:113107}). Their calculations for the increased electron and hole wave function spatial overlap is in agreement with PL measurements in Ref.(\onlinecite{founta:171901}) and supports the importance of geometry and Coulomb interaction for accurate modeling of the optical properties of a-plane GaN QDs. This manuscript is intended to provide theoretical and computational details of the approach the authors have used in Ref.(\onlinecite{schuh:092103}), in which parts of the results have already been published. Moreover the present manuscript discusses additional findings.

In the first part of the paper we present the geometry, shape and composition of the systems. Subsequently, we provide the theoretical background on the methodology applied in our approach and then end with the discussion of the results. To begin with, the approach to approximately determine the number of bound states within QDs is presented and it is explained how a calculation of an effective confinement potential is seperately possible for electrons and holes. The two-dimensional, single-particle effective potential takes into account the geometry, intrinsic fields and confinement due to the material system. Based on this quantity, an understanding of the wave functions's localization will be gained. In Sec. \ref{sec:results} we consider Coulomb interaction between the carriers in order to study the modification of carrier localization due to electron-hole attraction. Furthermore, we determine the critical interaction strength where the Coulomb interaction becomes dominant over the intrinsic elctrostatic fields.

\section{Geometry}
In this paper we present a theoretical analysis of pure lens-shaped InN/GaN QDs on a wetting-layer with a thickness of two lattice constants in the polar (c-plane) and nonpolar (m-plane) growth direction.
For the polar QD, we chose a geometry according to recent experiments, Refs.(\onlinecite{Rosenauer20111316,PhysRevB.83.115316,PSSB:PSSB201147143}). In order to separate the effect of the built-in fields from modifications of the QD geometry, we use the same shape and QD dimensions for the non-polar QD. Note that these properties of the QDs may change if the heterostructure is grown in a different crystal orientation, but we did not include these geometrical effects in the study in order to focus on the importance of Coulomb interaction solely as done in Ref.(\onlinecite{schuh:092103}). Nevertheless, due to the different crystal orientations and the discrete spatial lattice in our theory, the exact boundaries of the simulated nanostructure still differ within 10\% between both geometries. The dimensions of the polar (nonpolar) QD are $\Delta x\approx7.74$ nm ($\Delta x\approx2.76$ nm), $\Delta y\approx7.66$ nm ($\Delta y\approx7.66$ nm) and $\Delta z\approx3.11$ nm ($\Delta z\approx7.26$ nm).
In addition, the wetting-layer without the QD is also modelled in a seperate treatment for comparison. For the polar orientation we use a hexagonal supercell in combination with periodic boundary conditions to close the system and for the nonpolar orientation a cuboid one respectively. Throughout the paper we neglect spin-orbit splitting, as it is weak for InN with $\Delta_{so}=5$ meV, Ref.(\onlinecite{vurgaftman:3675}).

\section{Theory}
\subsection{Electronic properties}
For the calculation of the electronic properties we use an empirical tight-binding model (ETBM), because it allows for the possibility of a microscopic description of nanostructures like QDs and can be applied to larger systems. This is mainly due a spatial discretization on the scale of lattice sites in combination with the use of a minimal basis set of orbitals describing the relevant chemical bonds. In the past, ETBMs have been successfully used to describe the electronic states of bulk semiconductors and low-dimensional heterostructures such as quantum wells, nanorods and QDs., see e.g. Refs. (\onlinecite{PhysRev.94.1498,PhysRevB.50.5429,PhysRevB.68.235311,PhysRevB.72.165317,Korkusinski2008318}). In the following, a brief summary of the approach used in this paper is provided.
Our starting points are Loewdin-orthogonalized, effective local orbitals $\vert\mathbf{R},\alpha\rangle$, localized on the Bravais lattice at position $\mathbf{R}$ with a combined spin and orbital index $\alpha$ which are Wannier-like functions $w(\mathbf{r}-\mathbf{R})$. This type of ETBM is also referred to as an effective bond orbital model (EBOM), since it lacks the full atomistic resolution, but on the other hand there is some freedom to choose the tight binding basis as only hopping matrix elements $E^{\alpha,\alpha^{\prime}}_{\mathbf{R},\mathbf{R}^{\prime}}$ between these basis functions are needed. In principle, one can conserve the microscopic symmetry by extending the orbital basis as done in Ref.(\onlinecite{PhysRevB.68.235319}). In general, one has to perform a numerical fit or derive analytic expressions for $E^{\alpha,\alpha^{\prime}}_{\mathbf{R},\mathbf{R}^{\prime}}$ in terms of known physical parameters (effective masses, Luttinger parameters, bandgaps, critical points' energies, Kane parameters, etc.) either known from experiments or ab-initio calculations. The whole task has already been performed in Ref.(\onlinecite{PhysRevB.81.165316}) for the hexagonal Bravais lattice by using the $sp^{3}$-basis and taking into account matrix elements up to second nearest neighbors in combination with input parameters known from $G_{0}W_{0}$ calculations for wurtzite-nitride systems, for further details see Ref.(\onlinecite{1367-2630-7-1-126,rinke:161919,PhysRevB.77.075202,0268-1242-26-1-014037}). The dispersion $\varepsilon(\mathbf{k})$ throughout the whole Brillouin zone obtained within this model provides a quite reasonable description of the bulk electronic properties, since the conduction and three valence bands are pinned at all critical point energies except the $K$-point to ab-initio results while in addition the anisotropic effective masses at the $\Gamma$-point are included as well as the Kane-parameter. More details for the quality of the analytical parametrization can be found in Ref.(\onlinecite{PhysRevB.81.165316}).We proceed with the respective equations in a compact form.

In order to model a bulk system we start by defining an eigenfunction $\vert\mathbf{k}\rangle^{\text{Bulk}}$ as a linear combination with unknown coefficients $c_{\alpha}(\mathbf{k})$ of Bloch sums $\vert\mathbf{k},\alpha\rangle$ given by the Fourier sum of the ETBM basis $\vert\mathbf{R},\alpha\rangle$:
\begin{equation}
\vert\mathbf{k}\rangle^{\text{Bulk}}=\sum_{\alpha}c_{\alpha}(\mathbf{k})\underbrace{\frac{1}{\sqrt{N}}\sum_{\mathbf{R}}e^{i\mathbf{k}\mathbf{R}}\vert\mathbf{R},\alpha\rangle}_{\vert\mathbf{k},\alpha\rangle}.
\label{eq:psi_bulk}
\end{equation}
The corresponding matrix eigenvalue problem is of the form
\begin{equation}
\hat{H}^{\text{Bulk}}(\mathbf{k})=\frac{1}{N}\sum_{\mathbf{R},\mathbf{R}^{\prime}}e^{i\mathbf{k}(\mathbf{R}-\mathbf{R}^{\prime})}\underbrace{\langle\mathbf{R}^{\prime},\alpha^{\prime}\vert\hat{H}\vert\mathbf{R},\alpha\rangle}_{E^{\alpha,\alpha^{\prime}}_{\mathbf{R},\mathbf{R}^{\prime}}}
\label{eq:H_bulk}
\end{equation}
and provides the dispersion $\varepsilon(\mathbf{k})$ and coefficients $c_{\alpha}(\mathbf{k})$ after exact numerical diagonalization of Eq.(\ref{eq:H_bulk}). $N$ denotes the number of unit cells and ensures proper normalization. The reader is again referred to Ref.(\onlinecite{PhysRevB.81.165316}) for the details of this procedure.

The next step to generalize Eq.(\ref{eq:H_bulk}) for the description of a two-dimensional translational invariant system, i.e. a wetting-layer, is done by restricting the Fourier sum in Eq.(\ref{eq:H_bulk}) to the translational invariant in-plane dimensions solely. This approach corresponds to mapping the problem to a one-dimensional column whose effective real-space hopping matrix-elements are given by a partial Fourier sum, or simply speaking, we factorize the problem into an in-plane $(\perp)$ and out-of-plane ($\parallel$) contribution. This leads to the following equation in which $N_{\perp}$ is the number of in-plane unit cells:

\begin{eqnarray}
\hat{H}^{\text{Wl}}(\mathbf{k}_{\perp})=\frac{1}{N_{\perp}}\sum_{\mathbf{R}_{\perp},\mathbf{R}^{\prime}_{\perp}}e^{i\mathbf{k}_{\perp}(\mathbf{R}_{\perp}-\mathbf{R}^{\prime}_{\perp})}\nonumber\\
\langle\mathbf{R}^{\prime}_{\perp}+\mathbf{R}^{\prime}_{\parallel},\alpha^{\prime}\vert\hat{H}\vert\mathbf{R}_{\perp}+\mathbf{R}_{\parallel},\alpha\rangle.
\label{eq:H_wl}
\end{eqnarray}

There are additional matrix elements between different sites $\mathbf{R}_{\parallel}$, $\mathbf{R}_{\parallel}^{\prime}$ and not only between different orbitals $\alpha$, $\alpha^\prime$ compared to Eq.(\ref{eq:H_bulk}). Diagonalization of this matrix eigenvalue problem provides the subband dispersion $\varepsilon(\mathbf{k}_{\perp})$ and the coefficients $c_{\mathbf{R}_{\parallel},\alpha}(\mathbf{k}_{\perp})$ of the respective eigenstates:

\begin{equation}
\vert\mathbf{k}_{\perp}\rangle^{\text{Wl}}=\sum_{\mathbf{R}_{\parallel},\alpha}c_{\mathbf{R}_{\parallel},\alpha}(\mathbf{k}_{\perp})\underbrace{\frac{1}{\sqrt{N_{\perp}}}\sum_{\mathbf{R}_{\perp}}e^{i\mathbf{k}_{\perp}\mathbf{R}_{\perp}}\vert\mathbf{R}_{\perp}+\mathbf{R}_{\parallel},\alpha\rangle}_{\vert\mathbf{k}_{\perp},\mathbf{R}_{\parallel},\alpha\rangle}.
\label{eq:psi_wl}
\end{equation}

Setting up the Hamiltonian for a fully non-translational invariant system, i.e. a quantum dot, is now straightforward and yields:
\begin{equation}
\hat{H}^{\text{Qd}}=\langle\mathbf{R}^{\prime},\alpha^{\prime}\vert\hat{H}\vert\mathbf{R},\alpha\rangle.
\label{eq:H_qd}
\end{equation}
The eigenfunctions $\vert\psi\rangle^{\text{Qd}}$ are linear combinations of the ETBM basis with coefficients $c_{\mathbf{R},\alpha}$:
\begin{equation}
\vert\psi\rangle^{\text{Qd}}=\sum_{\mathbf{R},\alpha}c_{\mathbf{R},\alpha}\vert\mathbf{R},\alpha\rangle.
\label{eq:psi_qd}
\end{equation}

At the interfaces of two different materials $A/B$ of a heterostructure we use the average hopping of the two constituents. Furthermore an appropriate relative valence band offset $\Delta E_{V}$ for the on-site matrix elements of the two materials has to be incorporated in order to correctly model a heterostructure. In this work we use the valence band offset recommended in Ref.(\onlinecite{vurgaftman:3675}) which is 0.5 eV for the InN/GaN material system. The numerical diagonalization of Eq.(\ref{eq:H_qd}) is performed using the folded spectrum method of Ref.(\onlinecite{wang:2394}).

\subsection{Strain and elastic properties}
The elastic properties of the QDs under consideration have
been computed using a plane-wave based implementation of second-order
continuum elasticity theory as described in Ref.(\onlinecite{Marquardt2010765}).
Within the wurtzite crystal structure, the elastic energy
\begin{eqnarray}\label{eq:elastEn}
F&=&\frac{1}{2}\int_{V}d^3\mathbf{r} C_{11}(\epsilon_{xx}^2+\epsilon_{yy}^2)
+ C_{33}\epsilon_{zz}^2 \nonumber\\
&+& 2C_{12}\epsilon_{xx}\epsilon_{yy}+ 2C_{13}\epsilon_{zz}(\epsilon_{xx} + \epsilon_{yy})\nonumber\\
&+& 4C_{44}(\epsilon_{xz}^2 + \epsilon_{yz}^2)
+ 2(C_{11}-C_{12})\epsilon_{xy}^2
\end{eqnarray}
is minimized with respect to the displacements $\mathbf{u}(\mathbf{r})$.
Here the $C_{ij}=C_{ij}(\mathbf{r})$ are the elastic constants of the
wurtzite structure and V is the simulation cell volume, Ref.(\onlinecite{jogai:5050}). 
The strain tensor $\epsilon_{ij}(\mathbf{r})$ is related to the displacements
via:
\begin{equation}\label{eq:strainDisp}
\epsilon_{ij}(\mathbf{r}) = \frac{1}{2}\left(\frac{\partial u_i(\mathbf{r})}{\partial r_j}
+ \frac{\partial u_j(\mathbf{r})}{\partial r_i}\right) + \epsilon_{ij}^0(\mathbf{r})
\end{equation}
with the local intrinsic strain being defined by the lattice constants of the materials involved,
\begin{equation}\label{eq:intrStrain}
\epsilon_{ij}^0 = (\delta_{ij} - \delta_{iz}\delta_{jz})\frac{a_{\rm{ref}}-a(\mathbf{r})}{a(\mathbf{r})} 
+ \delta_{iz}\delta_{jz}\frac{c_{\rm{ref}}-c(\mathbf{r})}{c(\mathbf{r})} 
\end{equation}
where $a_{\rm{ref}}$ and $c_{\rm{ref}}$ are the reference lattice constants, chosen to be the bulk lattice
constants of the surrounding GaN.\\
Once the strain tensor $\epsilon_{ij}(\mathbf{r})$ is known, the built-in polarization potential
can be computed, arising from a piezoelectric and a spontaenous polarization. Within the wurtzite
lattice, the polarization is given by
\begin{equation}\label{eq:pol}
\mathbf{P}=\left(\begin{array}{c}
2e_{15}\epsilon_{xz}\\
2e_{15}\epsilon_{yz}\\
e_{31}(\epsilon_{xx} + \epsilon_{yy}) + e_{33}\epsilon_{zz}
\end{array}\right) + \left(\begin{array}{c}
0\\
0\\
P_{sp}
\end{array}\right).
\end{equation}
Here $e_{ij}=e_{ij}(\mathbf{r})$ denotes the piezoelectric constants and $P_{sp}$ is the spontaneous
polarization in the wurtzite structure. From the polarization vector $\mathbf{P}=\mathbf{P}(\mathbf{r})$, 
the polarization potential $\phi_P(\mathbf{r})$ is determined by solving the Poisson equation
\begin{equation}\label{eq:poisson}
\kappa_0\nabla[\kappa_r(\mathbf{r})\nabla \phi_P(\mathbf{r})]=\varrho_P(\mathbf{r}),
\end{equation}
where the polarization charge density is calculated as $\varrho_P(\mathbf{r})=-\nabla\mathbf{P}(\mathbf{r})$.
The influence of electromechanical coupling between strain and polarization potentials \cite{willatzen:024302}, was neglected since they are expected to be small. Indeed, only minor influences of this effect have been found in InGaN-based
nanostructures \cite{christmas:073522}. The polarization potential then enters the modelling of the electronic properties
of the QD as an additional potential contribution. For the calculation of the elastic and piezoelectric properties, the lattice, elastic and
piezoelectric constants from Ref.(\onlinecite{vurgaftman:3675}) have been employed. It is here important to
note that the choice of different piezoelectric constants (e.g. those from Ref. (\onlinecite{PhysRevB.79.075443}) can
induce significant modifications to the built-in electrostatic potential. In particular,
the choice of the parameter $e_{15}$ that has been controversely discussed recently
(\onlinecite{PSSC:PSSC200982609}) can lead to strong modifications mainly for the case of the nonpolar system.

The inclusion of strain and intrinsic fields into the calculation of electronic properties can be accomplished on different levels of sophistication. In principle, the hopping matrix elements $E^{\alpha,\alpha^{\prime}}_{\mathbf{R},\mathbf{R}^{\prime}}$ have to be altered due to the strain in a heterostructure: i) If the relaxed atomic positions which minimize the elastic energy of the nanostructure are known, one can use the new bond-lengths and angles to rescale the hopping matrix elements according to Harrison's law \cite{PhysRevB.20.2420}, though it might not be valid for our model. ii) In addition, the widely used two-center decomposition \cite{PhysRev.94.1498} allows for a direct analytical dependence of the hopping matrix elements as a function of bond-lengths and angles. iii) As a special case, the hamiltonian of the EBOM reproduces the hamiltonian of $\mathbf{k}\cdot\mathbf{p}$-theory in the limit of $\mathbf{k}=0$ by construction. Thus, it may be possible to derive an analytical parametrization for the $E^{\alpha,\alpha^{\prime}}_{\mathbf{R},\mathbf{R}^{\prime}}$ which directly includes a dependence on the deformation potentials \cite{0268-1242-26-1-014037} of the strain related $\mathbf{k}\cdot\mathbf{p}$ Hamiltonian\cite{PhysRevB.74.155322}.\\
A second important part arises from the intrinsic electrostatic potential energy,
\begin{equation}
\label{eq:elec_pot_energy}
V_{p}(\mathbf{r})=-e\cdot\phi_P(\mathbf{r}),
\end{equation}

which is naturally included as an on-site contribution to the tight-binding hamiltonian. This requires to first map and interpolate $V_{p}(\mathbf{r})$ obtained from continuum elastic theory to the tight-binding lattice $\mathbf{R}$ and then afterwards a symmetrization step. By rotating $V_{p}(\mathbf{R})$ six-times according to the $C_{6v}$ symmetry of the polar geometry (c-plane) and calculating a mean potential $\bar{V}_{p}(\mathbf{R})$, any artificial symmetry spoiling can be avoided. The same procedure is used for the nonpolar geometry (m-plane) with the difference that an inversion is used as a symmetry operation. In this work we only consider the contribution arising from $\bar{V}_{p}(\mathbf{R})$, since we assume it to be the more siginificant contribution compared to the modification of the hopping matrix elements.

\subsection{Interacting many-body problem}
The solution of the many-body problem is found by employing the full configuration interaction (FCI) framework, which is a method that has been successfully applied to QD systems with different basis sets \cite{PhysRevB.42.1713,PhysRevB.79.075443,PhysRevB.73.245327,PhysRevB.60.1819}. A main advantage over single-particle approximations like Hartree-Fock is, that all Coulomb correlations are included in the approach within the chosen basis.
The reader is referred to Ref.(\onlinecite{PhysRevB.61.7652}) for a more detailed discussion of this topic. The only limiting factor is the size of the single-particle basis, which restricts to low carrier densities. In our case, we are interested in studying the optical properties associated with one electron-hole pair, so there is only the electron-hole Coulomb interaction $V^{ehhe}$ present and scattering processes that change the number of carriers are not considered in our calculations. Furthermore, the small electron-hole exchange $V^{eheh}$ is neglected. The corresponding excited electron-hole (two-particle) eigenstate can be written as a linear combination of electron-hole basis states,
\begin{equation}
\vert\psi_{X}\rangle=\sum_{\alpha,\beta}c_{\alpha\beta}^{X}\hat{e}_{\alpha}^{\dagger}\hat{h}_{\beta}^{\dagger}\vert0\rangle,
\end{equation} 
where $\hat{e}_{\alpha}^{\dagger}$ and $\hat{h}_{\beta}^{\dagger}$ are electron and hole creation operators and $\vert0\rangle$ is the vacuum state.

In order to visualize the results of our calculations for the excitonic many-body eigenfunctions $\vert\psi_{X}\rangle$, it is convenient to independently provide the electron and hole part. Since a factorization of the density-operator
\begin{equation}
\hat{\rho}=\sum_{\alpha,\beta,\alpha^{\prime},\beta^{\prime}}\hat{e}_{\alpha}^{\dagger}\hat{h}_{\beta}^{\dagger}\vert0\rangle c_{\alpha\beta}^{X}c_{\alpha^{\prime}\beta^{\prime}}^{X*}\langle0\vert\hat{h}_{\beta^{\prime}}\hat{e}_{\alpha^{\prime}}.
\end{equation} is not possible, we take a partial trace over the electron or hole part in order to define a density-matrix for an electron (and corresponding for the hole) as
\begin{equation}
\hat{\rho}^{e}:=\sum_{\alpha,\alpha^{\prime}}\vert\alpha\rangle\underbrace{\sum_{\beta}c_{\alpha\beta}^{X}c_{\alpha^{\prime}\beta}^{X*}}_{\rho^{e}_{\alpha,\alpha^{\prime}}}\langle \alpha^{\prime}\vert,
\end{equation}
so that the probability density is finally given by
\begin{equation}
\langle\mathbf{R}\vert\hat{\rho}^{e}\vert\mathbf{R}\rangle=\sum_{\alpha,\alpha^{\prime}}\langle\mathbf{R}\vert\alpha\rangle\rho^{e}_{\alpha,\alpha^{\prime}}\langle\alpha^{\prime}\vert\mathbf{R}\rangle,
\label{mb_density}
\end{equation}
with $\vert\alpha\rangle=\hat{e}_{\alpha}^{\dagger}\vert0\rangle$ and $\vert\beta\rangle=\hat{h}_{\beta}^{\dagger}\vert0\rangle$.
The many-body Hamiltonian,
\begin{eqnarray}
	\hat{H}_X&=&\sum_\alpha E_\alpha\hat{e}^\dagger_\alpha\hat{e}_\alpha+\sum_\beta E_\beta\hat{h}^\dagger_\beta\hat{h}_\beta\\
	&-&\sum_{\alpha\beta\alpha^\prime\beta^\prime}(V_{\alpha\beta\beta^\prime\alpha^\prime}^{ehhe}\hat{e}^\dagger_\alpha\hat{e}_{\alpha^\prime}\hat{h}^\dagger_\beta\hat{h}_{\beta^\prime} +h.c.),
\end{eqnarray}
contains the single-particle energies $E_\alpha$, $E_\beta$ and the electron-hole Coulomb matrix elements $V_{\alpha\beta\beta^\prime\alpha^\prime}^{ehhe}$, which are evaluated in Fourier space,
\begin{eqnarray}
	V^{ehhe}_{\alpha\beta\beta'\alpha'}=\frac{1}{2V}\sum_{\vec{q}} \frac{e^2}{\kappa q^2}\left<\alpha\right| e^{i\vec{q}\vec{R}} \left|\alpha'\right>\left<\beta\right| e^{-i\vec{q}\vec{R}} \left|\beta'\right>
\end{eqnarray}
with the system volume $V$, elementary charge $e$ and permittivity $\kappa$. Alternatively, they can be calculated in real-space using the approximations in Refs.(\onlinecite{PhysRevB.73.245327}). If one restricts $\vert\alpha\rangle$ and $\vert\beta\rangle$ to a finite subspace of electron and hole single-particle states, which are localized in the QD region, a direct diagonalization of $\hat{H}_X$ becomes possible. In addition, the restriction to one electron hole pair allows for using a very large single-particle tight-binding eigenbasis entering the FCI calculation and in particular all bound QD states can be included this way.

\section{Results}
\label{sec:results}
\subsection{Subbandstructures}
\begin{figure}
\includegraphics[width=0.48\textwidth]{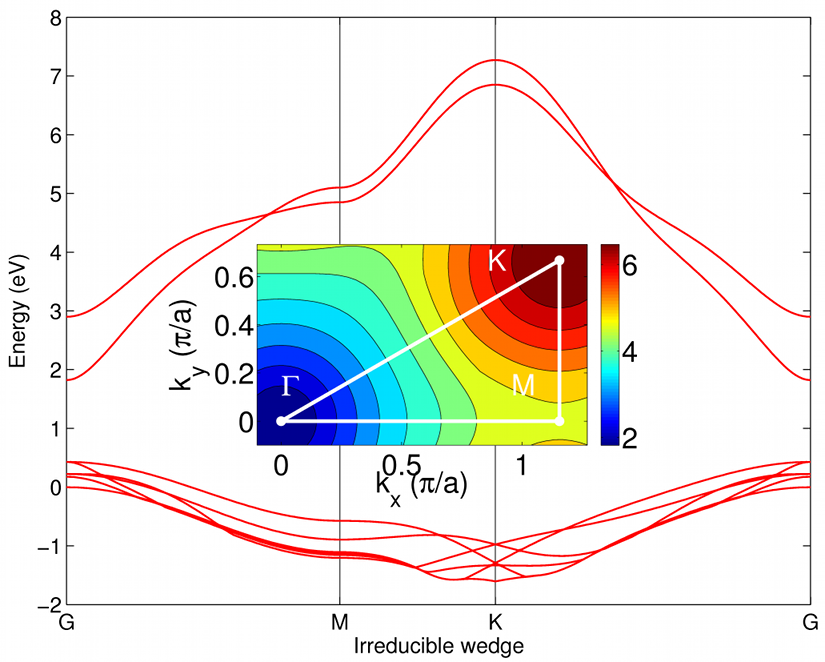}
\caption{\label{subbands_polar} Subbandstructure of the wetting-layer in the polar (c-plane) growth direction. The dispersion has hexagonal symmetry. As an inset the lowest two-dimensional conduction subband including the irreducible path is depicted.}
\end{figure}

\begin{figure}
\includegraphics[width=0.48\textwidth]{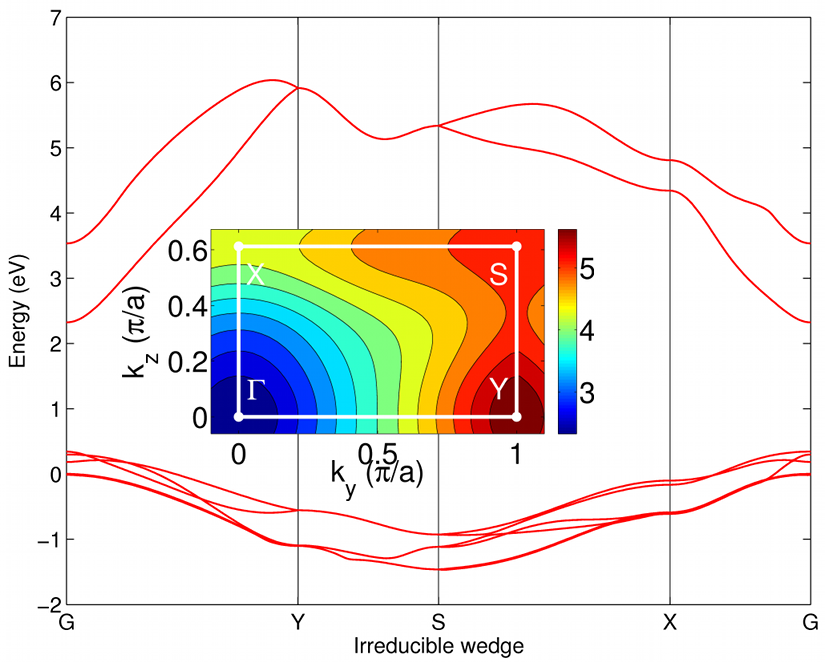}
\caption{\label{subbands_nonpolar} Subbandstructure of the wetting-layer in the nonpolar (m-plane) growth direction. The dispersion has primitive rectangular symmetry due to a change in the shape of the unit cell. As an inset the lowest two-dimensional conduction subband including the irreducible path is depicted.}
\end{figure}

One of the aims of this work is the investigation of excited states. For this purpose one has to know how many bound QD states exist in the given nanostructure. One solution is to calculate an energetic cutoff which separates all bound single-particle QD states from delocalized wetting-layer states. In order to determine this approximative energetic cutoff, which also limits the number of single-particle states considered as a basis for the FCI calculations later, we calculate $\epsilon(\mathbf{k}_{\perp})$ for the polar (c-plane) and the nonpolar (m-plane) wetting-layer (without the QD) using Eq.(\ref{eq:H_wl}).
We end up with 8 subbands in each case due to the sp$^3$ basis and the fact that the wetting-layer (WL) consists of two monolayers, as can be seen in Fig.(\ref{subbands_polar}) and in Fig.(\ref{subbands_nonpolar}) where the subbandstructures are depicted for their corresponding irreducible path. In the polar case, the initial sixfold symmetry of the Bravais-lattice is preserved while for the nonpolar wetting-layer the symmetry is reduced to a primitive rectangular one due to the fact that the structure has been rotated. The corresponding coordinates of the points of high symmetry are explicitely given in Tab.(\ref{symmetry_points}). To further illustrate this fact, the two-dimensional dispersion for the lowest conduction subband is plotted in Fig.(\ref{subbands_polar}) for the polar and Fig.(\ref{subbands_nonpolar}) for the nonpolar wetting-layer as an inset.
The energetic cutoff for the electrons is now estimated to be the minimum of the conduction subband dispersions, i.e. $min(\varepsilon^{CB}_{\mathbf{k}_{\perp}=0})$ and the maximum of the valence subband ones, i.e. $max(\varepsilon^{VB}_{\mathbf{k}_{\perp}=0})$ for the holes, respectively. This estimate is evaluated for both growth directions seperately and is justified, because if the QD is considered as only a small perturbation on an infinitely large wetting layer, these eigenenergies should in principle give the correct energetic limit where delocalized wetting-layer wave functions should start to appear.
In the polar (nonpolar) case we end up with an energetic cutoff of $\approx$1.821 (2.324) eV for the electrons and $\approx$0.429 (0.343) eV for the holes.

\begin{table}
\caption{\label{symmetry_points}Points of high symmetry for the polar (left) and nonpolar (right) wetting-layer with hexagonal ($C_{6v}$) and primitive rectangular ($C_{2v}$) symmetry respectively. Not given vector components $k_{i}$ are treated in real-space.}
\begin{ruledtabular}
\begin{tabular}{c||c|c|c||c|c|c|c}
 & $\Gamma$ & M & K & $\Gamma$ & Y & S & X\\
 \hline
 $k_x$ & 0 &  $2\pi/\sqrt{3}a$ & $2\pi/\sqrt{3}a$ & - & - & - & -\\
 $k_y$ & 0 & 0 & $2\pi/3a$ & 0 & $\pi/a$ & $\pi/a$ & 0\\
 $k_z$ & - & - & - & 0 & 0 & $\pi/c$ & $\pi/c$\\
\end{tabular}
\end{ruledtabular}
\end{table}

\subsection{Effective Confinement}
\begin{figure}
\includegraphics[width=0.48\textwidth]{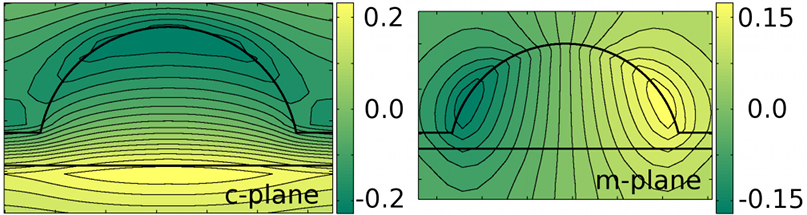}
\caption{\label{pol_fields} Intrinsic electrostatic potential energy in eV for the polar (c-plane) and nonpolar (m-plane) quantum dot orientation. Please note, that the actual numerical cell was much larger ($\approx23.1\times23.1\times9.3$ nm$^3$) as only the relevant excerpt is visualized.}
\end{figure}

We start by presenting an excerpt of the calculated intrinsic electrostatic potential energy in Fig.(\ref{pol_fields}) according to Eq.(\ref{eq:elec_pot_energy}) obtained by the methods outlined in Sec. 3.B, as it is required to calculate the effective confinement potential.
At first, there is a reduction of the maximum potential energy difference for the nonpolar orientation of about 100 meV from 488 meV to 384 meV compared to the polar case. This fact can be understood by the reduced spontaneous polarization part, since the surface area in the polar direction is also smaller. In each orientation, both fields tend to strongly seperate electron and hole wavefunctions into different directions. While in the polar case the electronic wavefunctions favor a localization in the top of the QD and the hole ones in the bottom near the wetting-layer, in the nonpolar geometry, both carriers should prefer the opposite sides of the QD lens. The above argumentation assumes ground-state wave functions, but related to this assumption two questions arise: Does this also apply for excited states and, what is the role of many-body effects ?

To address the first point, having calculated the intrinsic electrostatic potential energy $V_p(\mathbf{R})=-e\cdot\phi(\mathbf{R})$, it is now possible to determine an effective confinement potential by adding $\bar{V}_p(\mathbf{R})$ as a diagonal contribution to Eq.(\ref{eq:H_wl}). The effective confinement allows a first estimate of the energetic positions of the bound QD states in the heterostructure and provides more information on spatial localization perpendicular to the growth direction.

Obtaining this field is achieved by calculating the electronic dispersion according to Eq.(\ref{eq:H_wl}) for every spatial in-plane position $\mathbf{R}_{\perp}$ of the full three-dimensional problem, i.e. the quantum dot on the wetting-layer, but by using the actual material composition along $\mathbf{R}_{\parallel}$ in real space and by adding the position dependant $V_p(\mathbf{R}_\parallel)$ as an additional diagonal contribution. This procedure corresponds to a solution of the full three-dimensional problem by replacing the potential variation perpendicular to the growth direction by a translational invariant potential, but solving the Hamiltonian for each position seperately. The effective confinement for electrons and holes in the QD can now be approximated by plotting the electronic and hole ground state energy of the WL problem as a function of position perpendicular to the growth direction after exact diagonalization of Eq.(\ref{eq:H_wl}).

\begin{figure}
\includegraphics[width=0.48\textwidth]{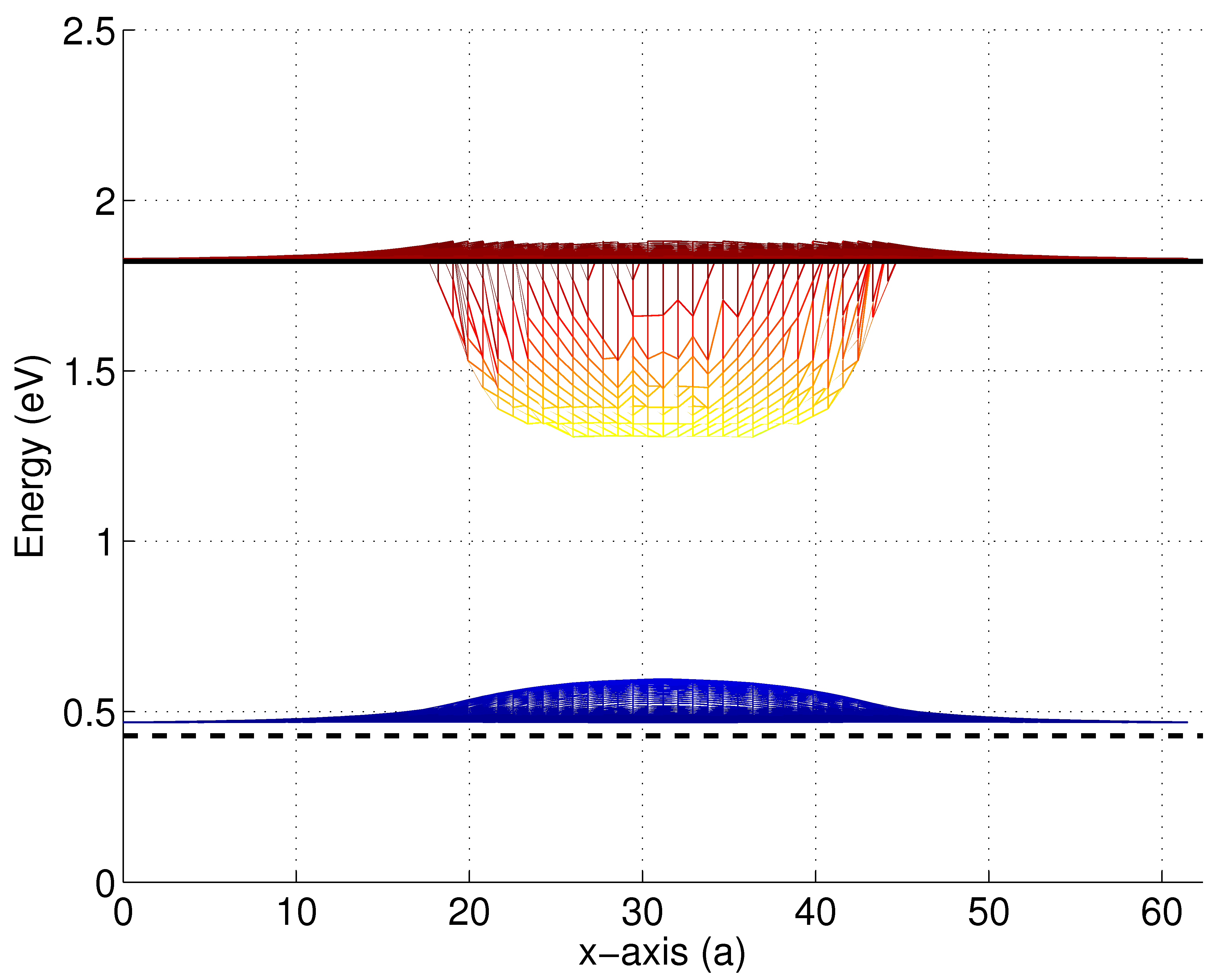}
\caption{\label{conf_polar} Side view of the effective confinement potential $V_{\text{eff}}(x)$ for the polar (c-plane) quantum dot for electrons (top) and holes (bottom). Each single line is a profile for different $y$-coordinates. The black solid line represents the electronic cutoff energy of $\approx$1.821 eV while the dashed one corresponds to the hole cutoff of $\approx$0.429 eV.}
\end{figure}

Let us begin by discussing the polar (c-plane) quantum dot case for which the effective confinement potential is depicted in Fig.(\ref{conf_polar}). One can expect the electronic wavefunctions with low energy to be localized in the top of the quantum dot and the holes in the bottom respectively. For both electrons and holes the potential seems to be of flat and quite homogeneous nature and we expect the ground states to be confined spatially in the QD. Excited states remain to be localized inside the QD and then gradually move to the wetting-layer. Remarkably, the effective electron potential is enhanced around the quantum dot for states with even higher energy than delocalized wetting-layer states, which is not the case for the holes as the electrostatic potential decays monotonically. The spatial variation of  $V_p(\mathbf{x,y})$, i.e. perpendicular to the growth direction, is symmetric as it should be since the quantum dot has a sixfold rotational symmetry. Thus, there is no region which favors or supports localization of carriers due to the lack of xy-anisotropy in the effective confinement inside the quantum dot. As a result, excited states can only gain overlap if they overcome the intrinsic fields along the $z$-direction.

\begin{figure}
\includegraphics[width=0.48\textwidth]{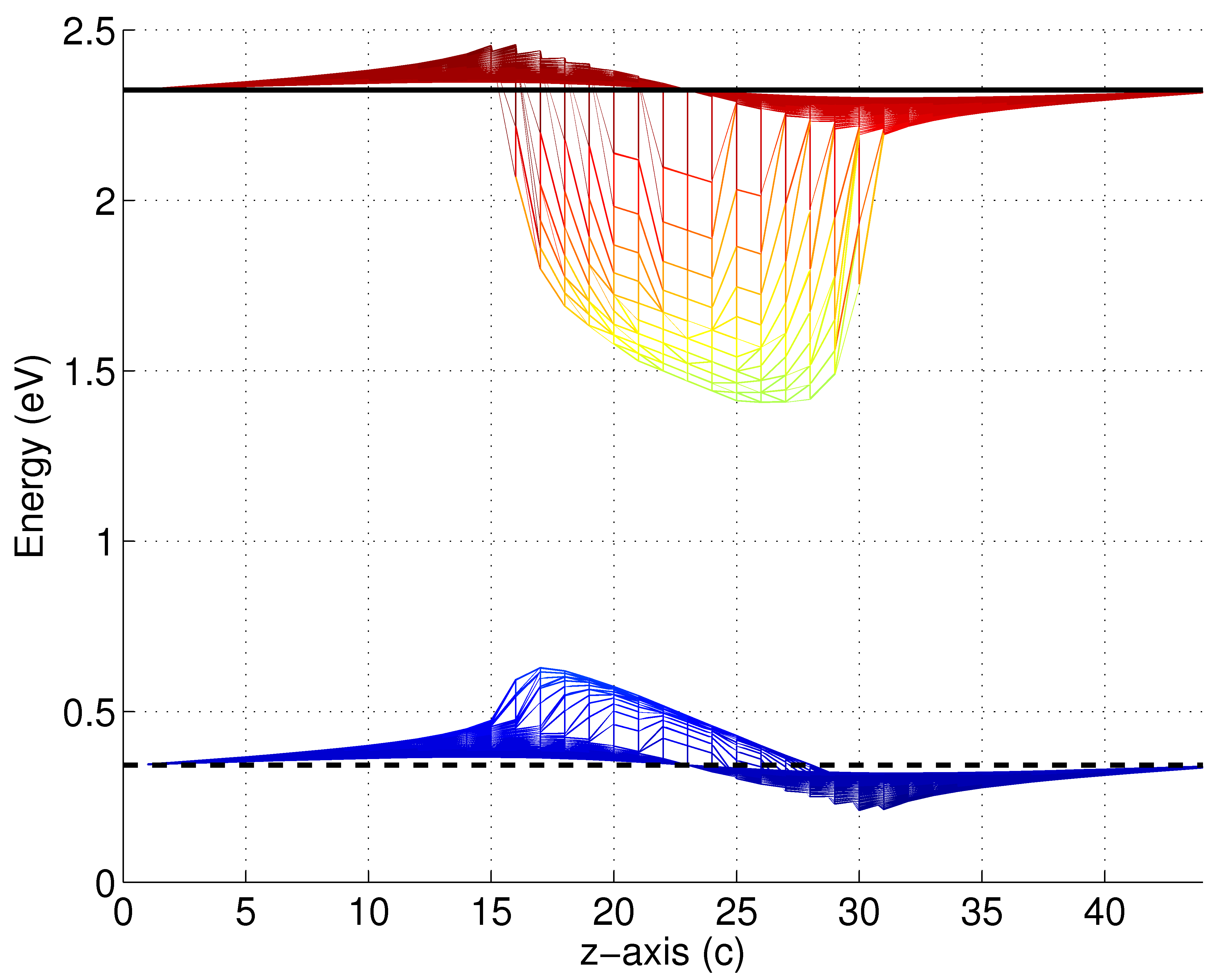}
\caption{\label{conf_nonpolar} Side view of the effective confinement potential $V_{\text{eff}}(z)$ for the nonpolar (m-plane) quantum dot for electrons (top) and holes (bottom). Each single line is a profile for different $y$-coordinates. The black solid line represents the electronic cutoff energy of $\approx$2.324 eV while the dashed one corresponds to the hole cutoff of $\approx$0.343 eV.}
\end{figure}

The results for the nonpolar geometry depicted in Fig.(\ref{conf_nonpolar}) show the modification of the carrier confinement due to the intrinsic electrostatic fields caused by the change in orientation. Generally, the bound hole states are localized in the lower part ($z<z_0$) of the QD, while the electronic wave functions are pushed to the upper part ($z> z_0$). Thus, electron and hole wavefunctions are separated for low energies and only excited states with a sufficient energy allow an increased overlap in the center of the QD. This is caused by the sloped carrier confinement which cleary shows the yz-anisotropy of $V_p(\mathbf{y,z})$, so that excited states can gain more overlap by moving perpendicularly to the growth direction opposed to the polar case. Furthermore, one can observe that because of this anisotropy the wetting-layer starts to become energetically more favorable than parts the QD if the energy is sufficiently high. The overall energetic position of the effective confinement potential $V_{\text{eff}}(\mathbf{R})$ is shifted to a deeper position in the polar case with more strongly bound states. This behavior is expected as the field strength is enhanced in the polar orientation.

As an additional benchmark for the energetic cutoffs calculated earlier, one can of course compare them to the present calculations including the intrinsic electrostatic potential.
The energetic cutoffs in the polar case are plotted as the black solid line for the electrons ($\approx$1.821 eV) and as a dashed one for the holes ($\approx$0.429 eV) in Fig.(\ref{conf_polar}). In the case of the electrons, the approximative value obtained by neglecting the intrinsic fields and the quantum dot is by 6.7 meV lower than the value obtained with this more accurate calculation. In contrast, the hole cutoff is underestimated by about 39.1 meV which can be observed from Fig.(\ref{conf_polar}) as it does not approach the potential in the limit of no QD. This fact indicates that the hole confinement is more sensitive to intrinsic fields compared to electrons, which is surprising, since one would expect a more remarkable deviation for both cases if the fields and quantum dot are neglected. The error introduced by this approximation only results in taking into account more single-particle hole states for the many-body calculation as actually to consider according to our approximative cutoff criteria. Note, that there is still some freedom in choosing the number of states, because the complete influence of the QD itself is not fully covered in these calculations using the factorization in Eq.(\ref{eq:H_wl}).

Furthermore, the energetic cutoff in the nonpolar orientation denoted by the black solid line for the electrons ($\approx$2.324 eV) and the dashed one for the holes ($\approx$0.343 eV) is visualized in Fig.(\ref{conf_nonpolar}). In both cases the approximative value is by 1.2 meV lower than the value obtained by this more exact calculation and the shift is energetically symmetric for both carriers as expected. As can be seen, the approximative cutoff is a reasonable choice to distinguish between quantum dot and wetting-layer states as it reproduces almost exactly the correct energetic limit of having no quantum dot and no intrinsic fields.

\subsection{Single-particle quantum dot states}
\begin{figure}
\includegraphics[width=0.48\textwidth]{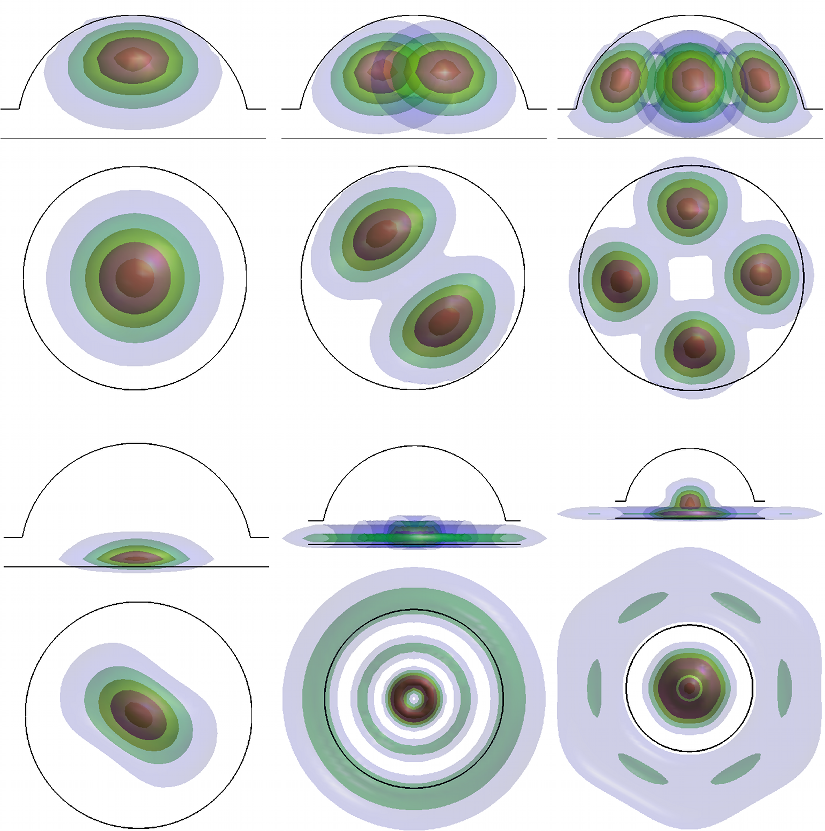}
\caption{\label{states_polar} Visualization of selected single-particle states by plotting isosurfaces of equal probability density in the polar (c-plane) geometry. The quantum dot is indicated by the black line and  the three lowest electrplotted from top and side view. In the upper part electronic wave functions are shown with increasing energy from left to right, i.e. the ground-state, an excited state in the middle and an highly excited state on the right. The same applies for the hole wave functions in the lower part.}
\end{figure}

\begin{figure}
\includegraphics[width=0.48\textwidth]{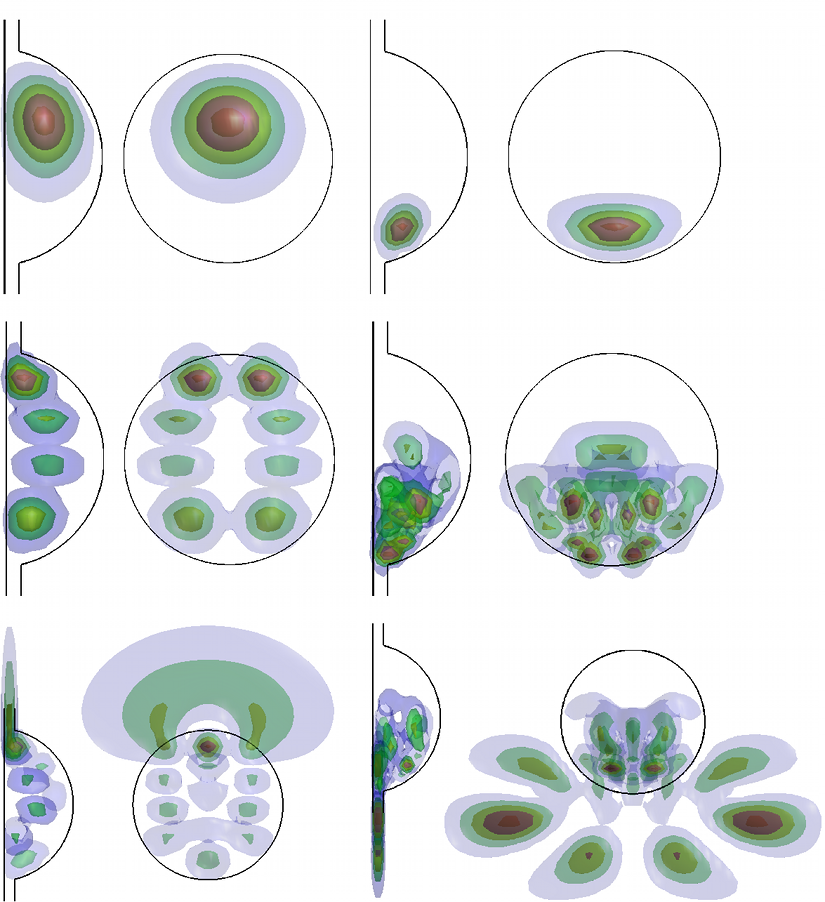}
\caption{\label{states_nonpolar} Visualization of some selected single-particle states by plotting isosurfaces of equal propability density in the nonpolar (m-plane) geometry. The quantum dot is indicated by the black line and each state is plotted from side- and topview. In the left part electronic wavefunctions are shown with increasing energy from top to bottom, i.e. the ground-state, an excited state and finally a highly excited state. The same applies for the hole wave functions in the right part.}
\end{figure}

In the following we support the conclusions drawn in the previous analysis by taking an explicit look at some exemplarily chosen single-particle wave functions. Representing a general type of wave function in terms of spatial localization, we have plotted some manually selected ones for the polar and nonpolar orientation respectively in Fig.(\ref{states_polar}) and Fig.(\ref{states_nonpolar}). In the polar case, all states obey the sixfold rotational symmetry of the underlying lattice. Furthermore, the electronic states, from the ground-state up to a highly excited state, are localized inside the quantum dot and move gradually from the top to the bottom. In contrast, the degenerate hole ground-states are deeply localized in the wetting-layer, while excited states tend to be more extended around the QD but remain inside the WL. Only the highly excited hole states exhibit a significant probability density inside the QD.

In contrast, the situation is completely different in the non-polar case as we can observe from Fig.(\ref{states_nonpolar}). At first, we are only left with a reflection as a symmetry operation with respect to the xz-plane at y=0 since the intrinsic electrostatic field with anisotropy in the z-direction spoils the former $C_{2v}$ symmetry of the problem. Furthermore, due to the sloped carrier confinement for electron and holes, their ground-state wave functions are strongly separated, even stronger than for the polar orientation due to the larger extent of the QD in the direction of the field. If we now increase the energy, excited states develop a finite probability density towards the center of the QD since the effective confinement now allows a localization with less separation. The higher excited electron states have a finite probability distribution throughout the whole QD region, while the hole states seem to be restricted to the lower half of the QD.
This can be understood by the fact that the effective hole mass is larger than the electron one, i.e. the holes are localized more strongly. Finally, we take a look at highly excited states which are supposed to feature localization inside the WL according to the discussion of the effective confinement. Exactly this is the case for both electrons and holes, which still have a non vanishing probability density inside the QD, but their major density is located right above and below the QD in combination with a large distribution over the WL plane.

\subsection{Coulomb interaction vs. QCSE}
\begin{figure}
\includegraphics[width=0.48\textwidth]{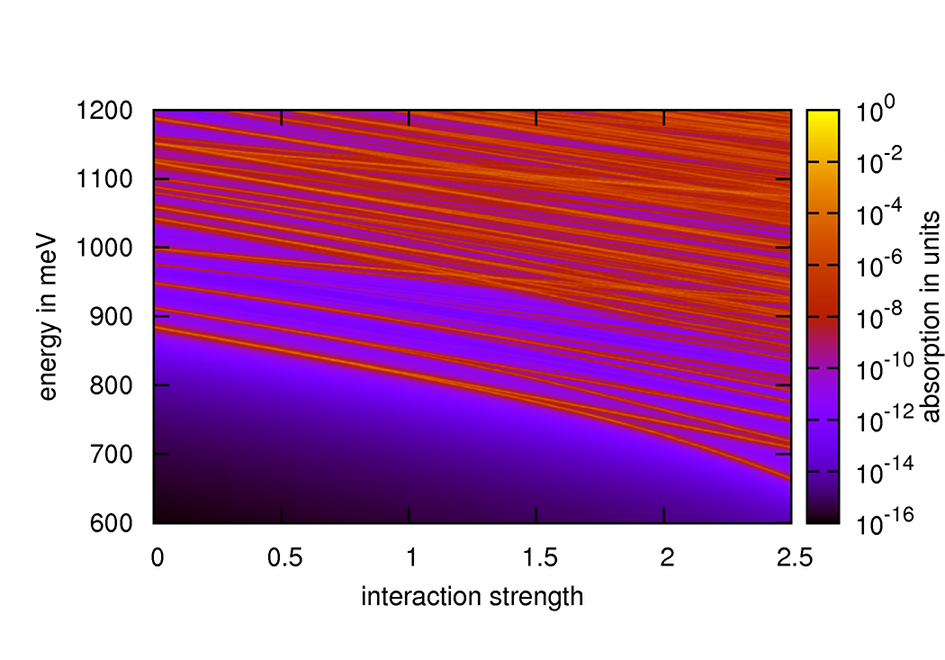}
\caption{\label{ww_c_polar} Absorption spectra of the polar quantum dot as a function of Coulomb interaction strength $\gamma$.}
\end{figure}

\begin{figure}
\includegraphics[width=0.48\textwidth]{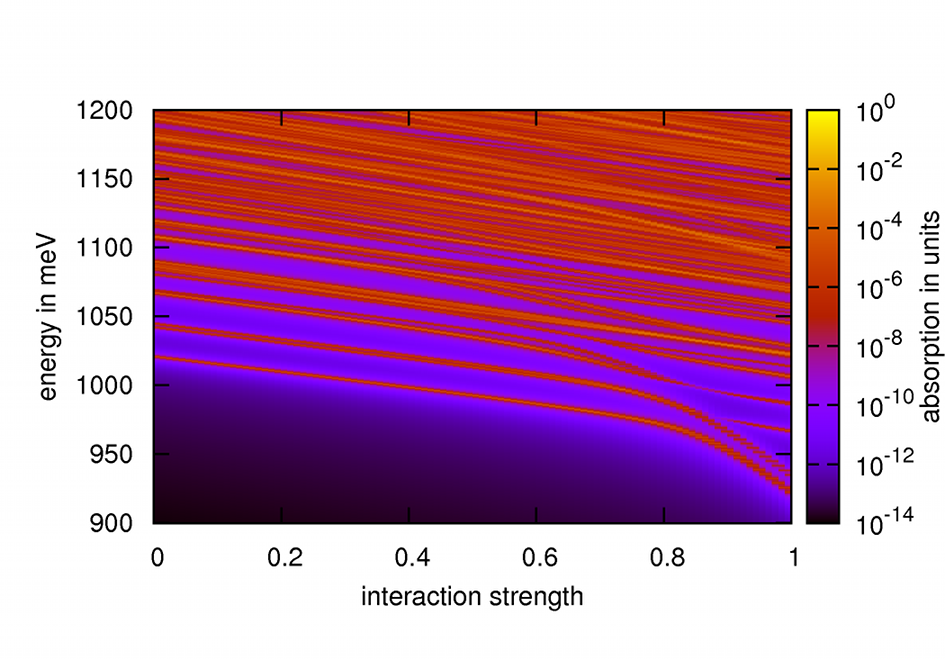}
\caption{\label{ww_c_nonpolar} Absorption spectra of the nonpolar quantum dot as a function of Coulomb interaction strength $\gamma$.}
\end{figure}

\begin{figure*}
\includegraphics[width=0.75\textwidth]{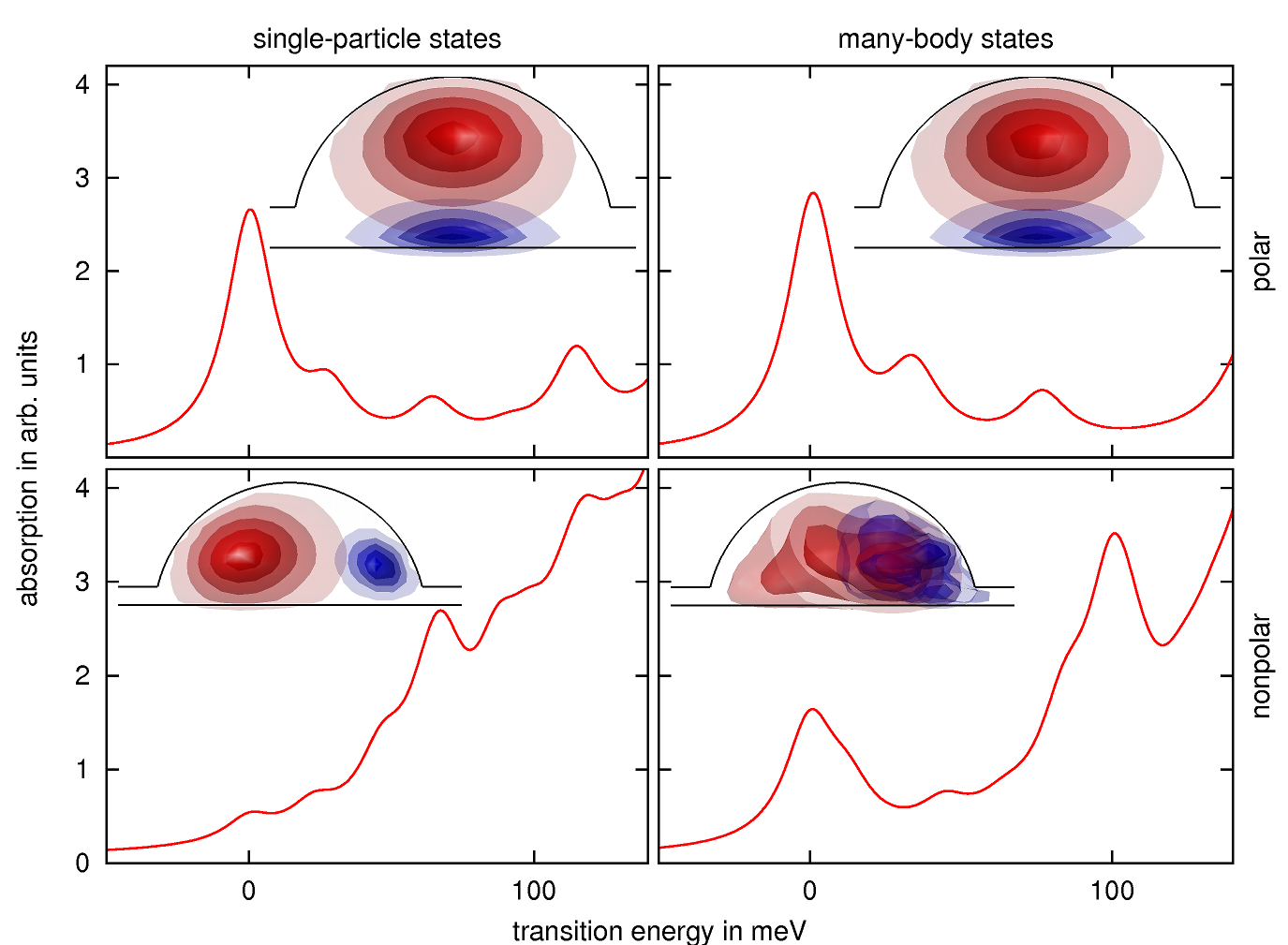}
\caption{\label{sp_vs_mb_states}Absorption spectra relative to the ground-state transition energy. Inset: Isosurfaces of equal probability density for electrons (red) and holes (blue) in the polar (upper panel) and nonpolar (lower panel) case including Coulomb interaction (right panel) compared to the single-particle description (left panel). The electronic part is localized in the top part of the quantum dot for the polar orientation and the left side in the nonpolar case. Please note, that the emission spectra (not shown) have been published in Ref.(\onlinecite{schuh:092103}).}
\end{figure*}

In this last section we extend the single-particle description of the system to the many-body treatment outlined earlier. The interplay between the QCSE and the Coulomb attraction manifests itself in the optical absorption spectra. In the following, linear optical spectra in connection with the excitation of one electron-hole pair are determined as a function of the parameter $\gamma$ for the strenghth of the Coulomb interaction, $\hat{H}_X=\hat{H}_0+\gamma\hat{H}_C$. The results of these calculations are depicted in Fig.(\ref{ww_c_polar}) and Fig.(\ref{ww_c_nonpolar}) and resemble the single-particle absorption spectra without Coulomb interaction in the case of $\gamma=0$, the true interaction strength for $\gamma=1$ and an unphysical artificial enhanced strength for $\gamma \ge 1$. For visualization purposes we have plotted the logarithm of the absorption.
For a system where Coulomb interaction only provides an energy renormalization, e.g. if it cannot compensate the intrinsic fields, and the electron and hole wave functions do not alter, we would expect that the energetic position of the emission line is reduced linearly with $\gamma$ if the interaction is increased. At the point where Coulomb interaction becomes dominant over the intrinsic fields and allows electrons and holes to gain energy by reducing their separation, i.e. they become spatially correlated, nonlinear effects should appear. Thus, the energetic position of the absorption lines should also be modified in a nonlinear way.
We note, that a very small Lorentzian broadening has to be used for a proper analysis of the effect.
We start by discussing the polar results in Fig.(\ref{ww_c_polar}) and observe that there are many emission lines with different slopes in the spectra due to different effective electron-hole pair distances. Nevertheless, the majority of these lines follow an almost linear trend for $\gamma \in [0,1]$. At an artificially enhanced Coulomb interaction of $\gamma\approx1.25$ several lines seem to split up into separate lines and nonlinear behavior is clearly identified for $\gamma\ge1.5$. This is due to the fact that new excitonic states form in the system and the intrinsic fields are compensated by the Coulomb interaction. Please note that new bright lines start to appear in the spectra due to an increased interaction strength, which lifts degeneracies and mixes bright single-particle states into former dark excitonic states.
In the nonpolar case depicted in Fig.(\ref{ww_c_nonpolar}) the situation is quite different, since most lines exhibit an identical linear slope. As can be seen, in the large energy regime a splitting of lines appears at about $\gamma\approx0.5-0.6$ and the ground state transition clearly follows a nonlinear behavior at $\gamma=0.8$. In addition there is a small indication of anticrossing of the absorption lines which also hints on strong energetic hybridization.
Summarizing both cases, we note that the unmodified Coulomb interaction is able to compensate the intrinsic fields in the nonpolar geometry and $\gamma=0.8$ is the critical value according to our calculations. In contrast, only an unphysically enhanced interaction strength of $\gamma\ge1.5$ can compensate the QCSE in the polar geometry. To further illustrate this fact, we have visualized the corresponding many-body and single-particle ground state eigenfunctions according to Eq.(\ref{mb_density}) for both geometries in Fig.(\ref{sp_vs_mb_states}) for the case of $\gamma=0$ and $\gamma=1$ including the corresponding absorption spectra relative to the ground-state transition energy. In contrast, in this figure a lorentzian broadening of 10 meV for elevated temperatures, Ref.(\onlinecite{PSSB:PSSB201147143}), has been included which simulates the effect of all inelastic scattering processes not explicitly taken into account.

As the Coulomb interaction is not strong enough to compensate the intrinsic electrostatic fields in the polar crystal orientation, there are only slight modifications of the excitonic ground-state wave function in comparison to the single-particle case. The main benefit is an energy gain due to the attractive electron-hole interaction. On the other hand in the nonpolar crystal orientation one can clearly observe that there is energy gain and also a strong spatial correlation of electrons and holes. As a result, the many-body ground-state wave function is modified drastically due to contributions of the formerly discussed excited single-particle states with increased overlap. Because the Coulomb interaction is strong enough for $\gamma\ge0.8$ to compensate the intrinsic fields, a new energetically more favorable ground-state is formed by mixing the excited singe-particle states. Thus, we end up with an enhanced absorption spectra in the nonpolar orientation. Additionally, there is an overall energetic shift of the ground-state absorption peak in both crystal orientations present.
The corresponding emission spectra (not shown) have been published in Ref.(\onlinecite{schuh:092103}).

\section{Summary}
In this work we investigated the interplay between Coulomb interaction and the QCSE in polar and nonpolar wurtzite InN/GaN lens-shaped quantum dots. This was achieved by combining continuum elasticity theory and a tight-binding model to describe elastic and single-particle electronic properties of these nanostructures properly. Before performing the full configuration interaction calculation for one electron-hole pair in order to solve the many-body problem, we calculated an effective confinement potential and discussed the single-particle states concerning their spatial localization inside the heterostructure on that basis. As a partial result, the sloped carrier confinement in the nonpolar geometry led to a different localization behaviour of excited states. In detail, electrons and holes tend to be less separated for excited states and surprisingly even higher excited states move spatially to the wetting-layer. The treatment of the interacting many-body problem of one exciton revealed, that the Coulomb interaction can compensate the intrinsic electrostatic fields and the critical relative strength is $\gamma=0.8$ in the nonpolar quantum dot. On the other hand, the Coulomb interaction is not strong enough to compete with the polarization fields in the polar quantum dot geometry, since only an artificially enhanced relative strength of $\gamma=1.5$ compensation took place. These insights were achieved identifying a nonlinear effect in the energies of optical transitions as a function of Coulomb interaction strength. Furthermore, the nonpolar ground-state absorption peak exhibits a quantitative enhancement due to Coulomb interaction and not only an energetic shift as in the case in the polar orientation. According to these findings, practical consequences for building blocks are related to higher carrier-densities, because in that regime one could expect multiexcitonic complexes in nonpolar structures to compensate the intrinsic fields much faster than in their polar counterparts. This could lead to an enhancement of optical recombination processes in that regime.

This work was supported by Deutsche Forschungsgemeinschaft under project no. Cz 31/14-1-3.

\bibliography{prb_refs}

\end{document}